\documentstyle[epsfig,natbib2,natbibmnfix]{mn}
\title[Ongoing Star Formation In AGN Host Galaxy Disk]
  {Ongoing Star Formation In AGN Host Galaxy Disks: A View From Core-collapse Supernovae}
\author[J. Wang et al.] {J.~Wang,$^1$\thanks{wj@bao.ac.cn}
  J. S.~Deng,$^1$ J. Y.~Wei$^1$
\newauthor\\ 
  $^1$ National Astronomical Observatories, Chinese Academy of Science}
\date{Released 2009 Xxxxx XX}

\pagerange{\pageref{firstpage}--\pageref{lastpage}} \pubyear{2009}

\def\LaTeX{L\kern-.36em\raise.3ex\hbox{a}\kern-.15em
    T\kern-.1667em\lower.7ex\hbox{E}\kern-.125emX}

\begin{document}

\label{firstpage}

\maketitle

\begin{abstract}
The normalized radial distribution of young stellar populations (and cold gas) in nearby 
galactic disks is compared between AGN host galaxies
and starforming galaxies (both with Hubble types between S0/a and Scd)
by using type II supernovae (SNe) as tracers. A subset of 140 SNe\,II with 
available supernova position measurements are selected from the SAI-SDSS 
image catalog by requiring available SDSS spectroscopy data of their host galaxies.  
Our sample is finally composed of 46 AGNs and 94 starforming galaxies.  
Both directly measured number distributions
and inferred surface density distributions indicate that 
a) the SNe detected in starforming galaxies follow an 
exponential law well; b) by contrast, the SNe detected in AGN host galaxies significantly deviate from an
exponential law, which is independent of both morphological type and redshift.
Specifically, we find a detection deficit around 
$R_{\mathrm{SN}}/R_{25,\mathrm{cor}}\sim0.5$ and an over-detection at outer 
region $R_{\mathrm{SN}}/R_{25,\mathrm{cor}}\sim0.6-0.8$.  
This finding provides a piece of evidence supporting that
there is a link between ongoing star formation (and cold gas reservoir) taking
place in the extended disk and central AGN activity. 

\end{abstract}


\section{Introduction}

It is now generally believed that active galactic nuclei (AGNs) play an
important role in galaxy formation and evolution. The growth of the central
supermassive black hole (SMBH) is suggested to be
related to the formation of the bulge of the host galaxy where the SMBH resides.
This evolutionary scenario is supported by the
well-established Magorrian relationship
(e.g., Magorrian et al. 1998; Tremaine et al. 2002; Ferrarese et al. 2006),
and by the fact that both star formation and AGN activity show similar
evolutions from z$\sim1$ to the current epoch (e.g., Ueda et al. 2003; 
Silverman et al. 2008).

So far, two kinds of mechanisms have been proposed to explain the co-evolution of
AGNs and their host galaxies. One possible mechanism is that both AGN activity and
formation of the bulge are triggered by a merger of two gas rich galaxies
(e.g., Granato et al. 2004; Springel et al. 2005; Hopkins et al. 2005).
Reichard et al. (2009) recently found
that more active AGNs with younger circumnuclear stellar populations are on average
associated with more lopsided host galaxies.
An alternative possibility is the gas inflow caused by the large scale
gravitational asymmetry of the host galaxies, such as a bar structure. 
Both mechanisms can produce an inflow of gas by transporting the angular momentum out of the gas.
The falling gas not only forms stars at the central region,
but also fuels the central SMBH.
The feedback of AGNs onto their host galaxies will likely regulate
the growth of the bulge by heating and expelling the surrounding gas through 
strong radio jets or other AGN-driven outflows (e.g., Croton et al. 2006; Hopkins \& Hernquist 2006;
Di Matteo et al. 2005).

The distribution of cold gas in AGN host galaxies is therefore crucial to the study of the co-evolution issue.
In addition to directly detecting the
gas distribution by the HI line emission, the gas distribution can be
approximately (and reasonably) substituted by the spatial distribution of young stellar
populations.
Although young stellar populations are frequently identified in
the host galaxies of some local AGNs (e.g., Cid Fernandes et al. 2001; Gonzalez Delgado et al. 2001;
Zhou et al. 2005; Wang et al. 2004; Wang \& Wei 2006; Mao et al. 2009),
their spatial distribution in the host galaxies is still poorly understood.
Combining the GALEX near-UV survey with the SDSS survey, Kauffmann et al. (2007) recently
found that in the local universe, although the AGN activity is strongly correlated with 
the age of the stars in the bulges (see also in Wang \& Wei 2008; Kewley et al. 2006; 
Wild et al. 2007), the most active AGNs are always associated with the bluest outer disks.
However, not all the galaxies with blue outer disks have an active AGN. 
This result therefore motivates the authors to believe that it could be understood if the 
amount of gas transported inward from disk is a variable.

In this paper, we investigate the co-evolution issue by
comparing the radial distribution of the core-collapse supernovae (cc-SNe) detected in AGN host galaxies
with the similar distribution of the cc-SNe detected in starforming galaxies.
Because cc-SNe are generally accepted to be produced by the explosion of massive stars
($\geq8-10M_\odot$) at the end of their lifetime $\la 10^{7.5}$yr (e.g., Woosley et al. 2002),
the radial distribution of cc-SNe reasonably represents the distribution
of young stellar populations. The advantage of this approach is that the result
does not strongly depend on the spatial resolution of the observations.


\section{Sample Selection and Analysis}

At first, the spectroscopic data from the SDSS Data Release 6 are cross-matched with
the SAI-SDSS image SNe catalog\footnote{The catalog is described in Prieto et al. (2008), and
can be downloaded from
http://www.astronomy.ohio-state.edu/\~\,prieto/snhosts/.}.
In our cross-matching, we require
that a) the angular offset of an individual SN measured from the center of the host galaxy
is less than 1\arcmin; b) the difference in redshift ($\Delta z$) between the SN and
corresponding host galaxy is less than 0.01.
The cross-matched sample in total contains 620 events, covering all three main
supernova types Ia, Ib/c and II. Among the sample, more than 97\% of the events show
$\Delta z<0.003$, which means that only for a few outliers,
the recessional velocity difference between the SNe and their host galaxies is
inferred to be larger than 900$\mathrm{km\ s^{-1}}$.
We further limit the sample to the SNe whose host galaxies have measured
photometric diameters and ratios of their minor and major axes.
To ensure adequate sample size and minimize the bias introduced by morphology types and host galaxy luminosity, 
we finally restrict our sample to the SNe whose a) host galaxies show late morphology types
ranging from S0/a to Scd (i.e., the T parameter is between 0 and 6.5), 
b) redshifts are less than 0.045.
Given that the SNe\,II are much more common than the SNe\,Ib/c,
only the SNe\,II are included in the analysis presented here.

The spectroscopic data taken by the SDSS are then analyzed to diagnose the central power source for
these galaxies.
For each narrow emission-line galaxy, the underlying
stellar absorption features are first removed from the observed spectrum by the principal
component analysis method (see Wang \& Wei 2008 for details).
The starlight-subtracted
spectrum is then used to measure emission line fluxes by the \it splot \rm task in the
IRAF\footnote{IRAF is distributed by the
National Optical Astronomical Observatories, which is operated by the Association
of Universities for Research in Astronomy, Inc., under cooperative agreement with the
National Science Foundation.} package. These emission-line galaxies are separated into
AGNs and starforming galaxies according to
the widely used BPT diagram (i.e., the [OIII]/H$\beta$ vs. [NII]/H$\alpha$
diagnostic diagram, Baldwin et al. 1981; Veilleux \& Osterbrock 1987).
The empirical demarcation line proposed by Kauffmann et al. (2003) is adopted in
the classification. Finally, 
there are in total 46 AGNs (hereafter AGN sample) and 94 starforming galaxies (hereafter 
SF sample) passing the above criteria. Note that 
the SNe\,II host AGNs are dominated
by type II AGNs that are on average a factor of 100 fainter in bolometric luminosity
than typical type I AGNs (see recent review in Ho 2008). The classifications of 
these galaxies are illustrated in Figure 1. AGNs and starforming galaxies are symbolized
by red-open squares and blue-solid squares, respectively. 
\begin{figure*}
 \includegraphics[width=150mm]{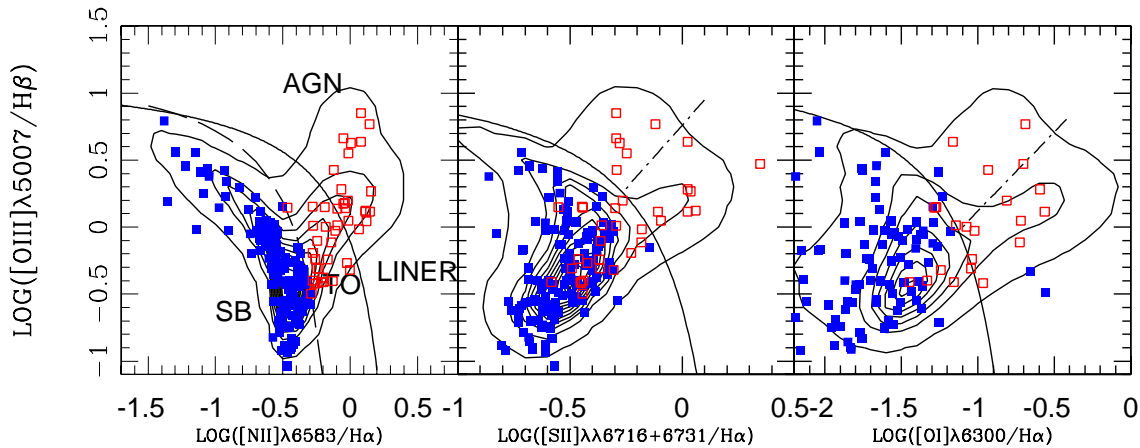}
 \vspace{0.1cm}
 \caption{The three diagnostic BPT diagrams for both AGN sample and SF sample. 
AGNs and starforming galaxies are presented by red-open squares and by blue-solid squares,
respectively. The solid lines show the theoretical
demarcation lines separating AGNs from star-forming galaxies proposed by
Kewley et al. (2001), and the long-dashed line the empirical line proposed
in Kauffmann et al. (2003), i.e., the demarcation line used in this paper. The underlying 
density contours are shown for a typical distribution of the narrow emission-line
galaxies taken from the MPA/JHU catalog (e.g., Kauffmann et al. 2003). Only the galaxies 
with S/N$>$20 and the emission lines detected with at least $3\sigma$ are considered.  
}
  \label{sample-figure}
\end{figure*}

Following previous studies (e.g., Petrosian \& Turatto 1990; Bartunov et al. 1992; van den Bergh 1997;
Petrosian et al. 2005),
the supernova relative distance ($R_{\mathrm{SN}}/R_{25,\mathrm{cor}}$) measured
from the center of the host galaxy is calculated for each SN by using the same method
used in Tsvetkov et al. (2004), where the projected distance of the SN from the center
of its host galaxy is $R_{\mathrm{SN}}=\sqrt{(\Delta\alpha)^2+(\Delta\delta)^2}$. 
At the direction along the position angle of the SN, 
$R_{25,\mathrm{cor}}$, the projected radius of the galaxy up to surface density 
of $25 \mathrm{mag\ arcsecond^{-2}}$ corrected for the inclination of the galaxy,
is calculated as 
\begin{equation}
R_{25,\mathrm{cor}}=\frac{d_{25}}{2\sqrt{\cos^2\theta+b^2\sin^2\theta}}
\end{equation}
where $d_{25}$ is the measured galaxy photometric diameter up to surface density of 
$25 \mathrm{mag\ arcsecond^{-2}}$, $\theta$ the position angle of an 
individual SN, and $b$ the ratio between major and minor axes.

\section{Supernovae in AGN Host Galaxies: A Significant 
Deviation From An Exponential Law}

Even though the two sub-samples are selected by taking the cuts in morphological 
type and redshift, the two sub-samples do not show identical distributions 
of their morphological types and redshifts. To alleviate the possible biases, 
we assign a weight for each galaxy in the SF sample. At first, 
the distributions of the morphological types are compared between the two sub-samples.
A weight is assigned to each galaxy in the SF sample by requiring the 
two distributions are identical. A second weight could be obtained through the 
similar procedure but for redshifts. In each procedure, we adopt a bin size that 
ensures each bin contains at least one object. The total weight associated with each
galaxy in the SF sample is derived by multiplying the two weights.

By taking the calculated weights into account, 
Figure 2 compares the histogram of the 
radial distribution of $R_{\mathrm{SN}}/R_{25,\mathrm{cor}}$ of the 
SNe\,II discovered in the AGN host galaxies with the similar plot of the SNe\,II discovered in the starforming
galaxies. The overplotted error-bar for each bin corresponds its 1$\sigma$ Poisson noise.
Both radial distributions show a deficit of
SNe within the region $R_{\mathrm{SN}}/R_{25,\mathrm{cor}}<0.2$ (see also in e.g., Petrosian et al. 2005; 
Hakobyan et al. 2009),
which is an observational bias, i.e., the Shaw (1979) effect. It is a challenging task to uncover
a supernova event from galactic centers, both because of the luminous background and
because of the possible heavy extinction in the galactic centers.

\begin{figure}
 \includegraphics[width=84mm]{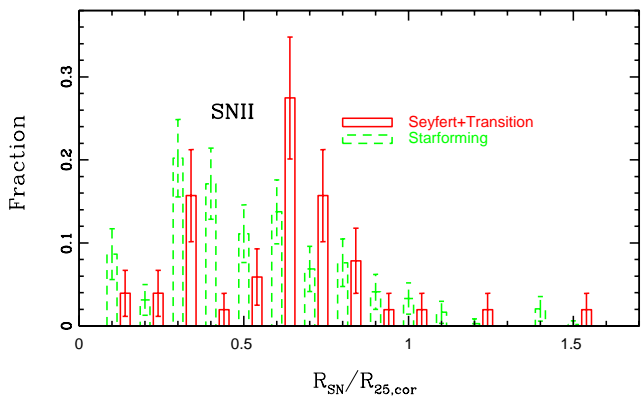}
 \vspace{0.1cm}
 \caption{The distribution of relative distance of the SNe detected in 
the AGN host galaxies (the histograms by red-solid line) is compared with 
the similar distribution of the SNe discovered in the starforming galaxies
(the histograms by green-dashed line). The error-bar overplotted for each bin
is the 1$\sigma$ value assuming a Poisson distribution. 
} 
  \label{sample-figure}
\end{figure}

One can see from the figure that, 
in the outer disk, the $R_{\mathrm{SN}}/R_{25,\mathrm{cor}}$ distribution
of the SNe\,II discovered in the starforming galaxies decreases smoothly with the radial distance
measured from the galactic centers,
which bears a strong similarity to the starlight distribution in the galactic disk.
Hakobyan et al. (2009) suggested that the distribution of the relative distance of 
cc-SNe from the centers of their host galaxies could be appropriately described by an exponential law. 
Compared with the case in the starforming galaxies,
the SNe\,II discovered in AGN host galaxies show a different radial distribution with
both an evident reduced fraction at $R_{\mathrm{SN}}/R_{25,\mathrm{cor}}\sim0.4$ and
a clear over-detection in the outer region within $R_{\mathrm{SN}}/R_{25,\mathrm{cor}}\sim0.6-0.8$,
i.e., a bimodal radial distribution. As shown in the figure, both features are significant 
at a confidence level no less than 1$\sigma$.
We further roughly quantify the peak of the over-detection to be at
$R_{\mathrm{SN}}/R_{25,\mathrm{cor}}\sim0.6$. By considering the objects with 
$R_{\mathrm{SN}}/R_{25,\mathrm{cor}}\leq1$, the 
Gehan's Generalized Wilcoxon two-sample statistical tests
show the two distributions are drawn from the same parent population at a confidence level of
4\% (for permutation variance) and 5\% (for hypergeometric variance). A two-side Kolmogorov-Smirnov
test is performed on both samples. The test yields a max discrepancy of 0.29 with a corresponding 
probability that the two samples match of 1.3\%.

It is noted that a similar bimodal distribution is also recently identified in
Hakobyan et al. (2009) who studied the relative radial distribution of the cc-SNe
from the Asiago catalog. By separating the sample into two groups, i.e., active- and non-active 
galaxies, our study indicates that the second peak shown in the Figure 4 of 
Hakobyan et al. (2009) is in fact mainly contributed by AGNs.

Figure 3 plots the relative distances of the 
SNe\,II calculated from the centers of their hosts vs. the morphological type of the host 
galaxies (the upper panel) and redshifts (the lower panel). The objects in the AGN sample
and in the SF sample are shown by red crosses and green circles, respectively.     
In both panels, starforming galaxies are continually distributed in the diagram. By 
contrast, the bimodal radial distribution can still be clearly identified for AGNs even when one 
compares the relative distances between AGNs and starforming galaxies at a given morphological type
or redshift. There is an obvious gap at $R_{\mathrm{SN}}/R_{25,\mathrm{cor}}\sim0.4-0.5$ separating
the AGNs into two sub-groups.  
This figure therefore strongly suggests that the bimodal distribution
revealed for AGNs is robust, i.e., not correlated to morphological type and luminosity 
of the supernova host galaxies. 

\begin{figure}
 \includegraphics[width=84mm]{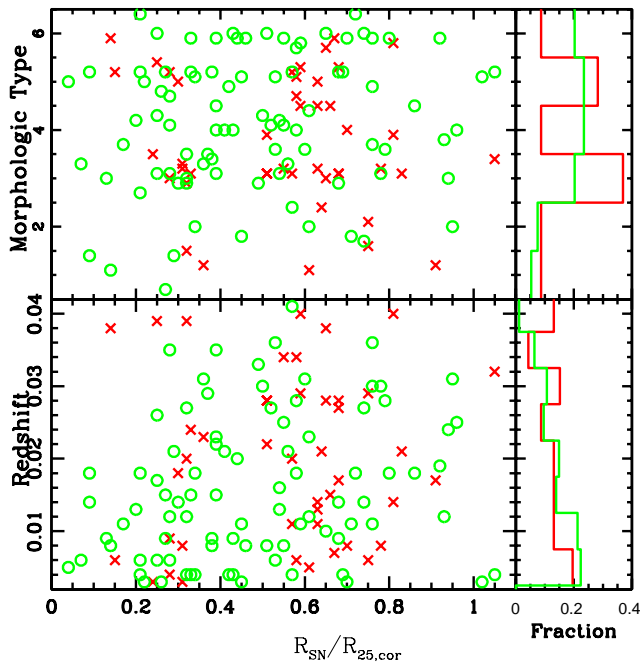}
 \vspace{0.1cm}
 \caption{\textit{upper-left panel:} morphological types plotted against 
the supernova relative distance $R_{\mathrm{SN}}/R_{25,\mathrm{cor}}$ 
measured from the centers of their host galaxies. The AGNs and starforming 
galaxies are shown by the red crosses and green open-circles, respectively.
\textit{lower-left panel:} the same as the upper panel, but for redshifts.
In both panels, the starforming galaxies are continually distributed, and 
the AGNs show a bimodal distribution that is independent of morphological
type and redshift. A gap at $R_{\mathrm{SN}}/R_{25,\mathrm{cor}}\sim0.4-0.5$
clearly separate the AGNs into two sub-groups. \textit{right panels:} histograms
of morphological type and redshift for both AGN- and starforming samples.  
}
  \label{sample-figure}
\end{figure}

As an additional test, the bimodal radial distribution of the SNe\,II discovered in 
AGN host galaxies is still significant if we examine 
the issue more physically. Figure 4 shows the surface density distribution of the SNe\,II as 
a function of $R_{\mathrm{SN}}/R_{25,\mathrm{cor}}$ for both AGN and SF samples.
The surface density is calculated as $\Sigma_{\mathrm{SN}}=N/S$ for each distance bin, 
where $S=2\pi r\Delta r$ is the 
area of a circle with a radius $r$ and a width $\Delta r$, and $N$ the number of 
SNe detected within the circle. The distributions plotted in Figure 1 are rebinned into a single 
bin for the outer region $R_{\mathrm{SN}}/R_{25,\mathrm{cor}}>1$. In Figure 4, 
the AGN and SF samples are symbolized by red-triangles and green-open-circles, respectively. 
The 1$\sigma$ Gaussian uncertainty over-plotted in the diagram is calculated according 
to the error tables given in Gehrels (1986). 
The surface density distribution of the SF sample 
is weighted through the same method described above.

The surface density of cc-SNe is usually well modelled as an exponential profile as a function 
of $R_{\mathrm{SN}}/R_{25,\mathrm{cor}}$ (e.g., Hakobyan et al. 2009; Barunov et al. 1992). 
Assuming an exponential model $\Sigma_{\mathrm{SN}}=\Sigma_{\mathrm{0,SN}}\exp(r/h)$, where
$r=R_{\mathrm{SN}}/R_{25,\mathrm{cor}}$ and $h$ is the length scale in units of $R_{25,\mathrm{cor}}$,
the green long-dashed line in Figure 4 plots the best fitting model for the SF sample. The 
two points with $R_{\mathrm{SN}}/R_{25,\mathrm{cor}}<0.2$ are excluded in the fitting because 
of the Shaw effect. The fitting yields a length scale $h=0.23\pm0.03 R_{25,\mathrm{cor}}$.
Our length scale is slightly less than the value obtained in Hakobyan et al. (2009, and references 
therein). The slight difference could possibly result from two causes. First,   
it is emphasized that the exponential model is obtained here from SF sample alone, which
differs from the previous studies. In these studies, the authors did not separate their samples into 
sub-groups according to the central engine of supernova host galaxies. In fact, our study 
indicates that the surface density of the SNe\,II discovered in AGN host galaxies deviates 
from an exponential profile significantly. Secondly, our sample is selected by requiring individual SN to
lie within 1 arcminute of the corresponding host galaxy center. These selection causes the 
sample to be biased against the SNe discovered at the edge of very nearby host galaxies. Note that 
the main conclusion presented in the current paper can not be affected by the bias since both 
AGN- and starforming sub-samples are selected by the same method.   

To illustrate the deviation from an exponential profile for the AGN sample, 
we vertically shift the best fitting 
derived for the SF sample by an amount of -0.31 dex ($=\log(47/95)$)
by fixing the exponential index.
The shifted exponential model is 
drawn by a red short-dashed line in Figure 4. The model obviously provides a good match for 
the points at the two ends (by excluding the two points with $R_{\mathrm{SN}}/R_{25,\mathrm{cor}}<0.2$ 
as well). Comparing the model with the calculated surface density allows us to identify an 
over-density at the region $R_{\mathrm{SN}}/R_{25,\mathrm{cor}}\sim0.6-0.8$ and a low density 
at $R_{\mathrm{SN}}/R_{25,\mathrm{cor}}\sim0.4$, 
which agrees with the analysis based upon 
the directly measured number distributions.

\begin{figure}

 \includegraphics[width=84mm]{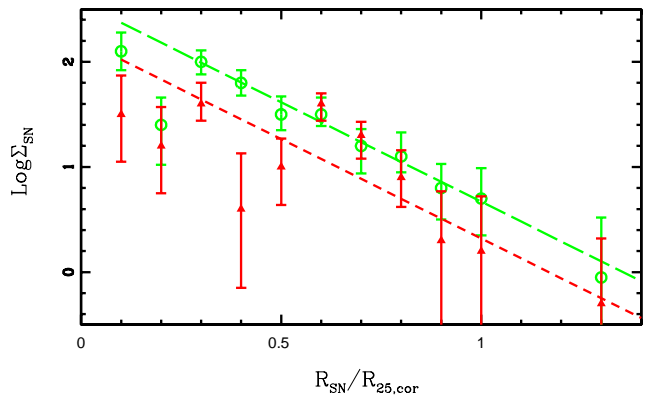}
 \vspace{0.1cm}
 \caption{A comparison of the two surface density distributions. The green open-circles 
present the SNe\,II detected in starforming galaxies, and the red triangles the 
SNe\,II detected in AGN host galaxies. The $1\sigma$ Gaussian error-bars overplotted are 
estimated from the error tables given in Gehrels (1986). 
The surface density distribution of the SNe\,II detected in the 
starforming galaxies can be well modelled as an exponential law shown by the 
green long-dashed line. To illustrate the deviation from an exponential law for 
the SNe\,II detected in AGN host galaxies, we vertically shift the previously 
modelled exponential law by an amount of -0.3 dex (see the red short-dashed line).} 
  \label{sample-figure}
\end{figure}

\section{Conclusion And Discussion}

Because SNe\,II are generated from the explosion of massive stars ($\geq8M_\odot$),
the SNe\,II radial distribution in their host galaxies reasonably reflects not only the radial
distribution of young stellar populations, but also the radial distribution of cold gas, assuming
a uniform supernova rate. By comparing the radial distributions of the SNe\,II detected in AGNs and
the similar distribution of the SNe\,II detected in  
starforming galaxies, we find that the supernova radial distribution in AGN host galaxies deviates greatly 
from the exponential model that can describe the radial distribution in starforming galaxies well. 
Both directly measured number distribution and inferred surface density indicate that SNe detected 
in AGN host galaxies show a bimodal distribution as a function of radius.   

The comparison of the radial distributions of the SNe\,II detected in the two types of supernova host galaxy
allows us to argue that the existence of the AGN activities is connected with the gas reservoir located
in the extended galactic disk, which agrees with
the previous studies. Kauffmann et al. (2007) identified a UV-light excess
in the extended disk for local AGN host galaxies. Hunt et al. (1999) suggested that the AGN host
galaxies show a larger gas fraction in their disk than non-active galaxies.
Using deep imaging from Spitzer and GALEX, Zheng et al. (2009) suggested that the
star formation in massive galaxies at $z<1$ mainly takes place in
the isolated disks.
Moreover, stellar rings in the disks are more frequently
identified in AGN host galaxies than in starforming galaxies (Hunt \& Malkan 1999).
By examining the spatially resolved stellar populations of 8 AGNs at $z\sim1$,
Ammons et al. (2009) arrived at a conclusion that the strong type II AGNs are
associated with extended star formation activities.

The star formation occurring in non-active galaxies and star formation
associated with SMBH accretion appear to be different events with different origin.
The radial distribution of the SNe\,II in the starforming galaxies can be 
well modelled by an exponential model, which means the gas distribution
in these galactic disks is not significantly disturbed.
Some particular dynamical mechanisms are necessary
in AGN host galaxies to redistribute gas to trigger both large-scale star
formation occurring in the outer disk and central SMBH activity (and also
associated circumnuclear star formation).

So far, several mechanisms have been proposed to link the central AGNs with the outer parts
of their host galaxies.
Kauffmann et al. (2007) proposed that
the gas distributed in the outer disk of AGN host galaxies could stem from
the accretion of gas from an external source. In addition, the redistribution of 
gas could be resulted from minor merger or major merger of two galaxies (e.g.,
Martini 2004 and references therein). Recent numerical simulations indicated that
the stars and gas could survive to re-form a disk in the merger of two gas rich
galaxies (e.g., Hopkins et al. 2009; Hammer et al. 2005). 
In the merger process, the gas within some characteristic radius loses its angular momentum quickly,
and sinks into the galactic central region by the gravitational attraction.
The gas that survives outside of the characteristic radius will descend to
form a new disk if the strong AGN feedback is taken into account.
The multiple disks 
produced by interactions are indeed observed in individual local Seyfert galaxies, 
e.g., Mark 315, a Seyfert 1.5 galaxy (Ciroi et al. 2005). Reichard et al. (2009)
recently reported a connection between AGN activity and lopsidedness of their host galaxies.
An outer loop and an arc with blue colors were observed in Seyfert 1.8 galaxy Mark 334 at 
an radius $r\sim$20-30\arcsec from the center (Smirnova \& Moissev 2009). The blue colors suggest
the existence of young stellar populations ($\sim$0.5-1 Gyr) that are formed in 
the interaction process. Adopting the characteristic radius of the galaxy $R_{25}\approx50$\arcsec 
estimated from the Figure 4 in Smirnova \& Moissev (2009), the relative distance 
of the outer loop and arc is inferred to be $\sim0.4-0.6$.

Besides the merger scenario, it is now generally believed that the bar-driven gas inflow is
related to the formation of the gas rings (Buta \& Combes 1996).
Theoretical and N-body simulation studies indicated that the gravitational asymmetry
caused by the bars transports gas angular momentum. The migration of the angular momentum
results in a gas inflow within the corotation radius and an outflow of gas out
of the corotation radius (e.g., Sellwood \& Wilkinson 1993; Athanassoula 2003).
The gas redistribution fuels central AGNs and circumnuclear starbursts, destroys the bars
(e.g., Bournaud \& Combes 2002), and regulates
the gas into a stable configuration (e.g., rings) by itself.

\section*{Acknowledgements}

We thank the anonymous referee for his/her valuable comments and suggestions that are 
helpful in improving the paper.  
The authors thank James Wicker for help with language.
This study
uses the SDSS archive data that was created and distributed by the Alfred P. Sloan
Foundation. This work was supported by the National Science Foundation of
China (under grants 10673014 and 10803008) and by the National Basic Research Program of
China (grant 2009CB824800).

\bibliographystyle{mn}

\end{document}